\documentclass{aa}
\usepackage{epsfig}
\usepackage{times}
\usepackage{mathptm}
\epsfclipon
\begin{document}

   \thesaurus{12     % Cosmology.
              (12.07.1;  % Gravitational lensing
               11.17.3;  % quasars: general
               11.17.4)} % quasars: individual: HE~0230$-$2130
   \title{The new complex gravitational lens system HE~0230$-$2130%
          \thanks{Partly based on observations collected at the European
                  Southern Observatory, La Silla, Chile.}
         }

   \author{L. Wisotzki\inst{1}
           \and
           N. Christlieb\inst{1}
           \and
           M. C. Liu\inst{2}\thanks{Visiting astronomers at the Cerro Tololo
                                 Interamerican Observatory, National 
                                 Optical Astronomy Observatories, which are
                                 operated by the Association of Universities
                                 for Research in Astronomy, Inc., under
                                 cooperative agreement with the National
                                 Science Foundation.}
           \and
           J. Maza\inst{3}
           \and
           N. D. Morgan\inst{4}$^{\star\star}$
           \and
           P.L. Schechter\inst{4}$^{\star\star}$
          }

   \offprints{L. Wisotzki, lwisotzki@hs.uni-hamburg.de}

   \institute{
              Hamburger Sternwarte, Universit\"at Hamburg, Gojenbergsweg 112, 
              21029 Hamburg, Germany
         \and
              Dept.\ of Astronomy, University of California, Berkeley,
              CA 94720, USA
         \and 
              Departamento Astronomia, Universidad de Chile, Casilla
              36-D, Santiago, Chile
         \and
              Dept.\ of Physics, Massachusets Institute of
              Technology, Cambridge, MA 02138, USA
             }

   \date{Received; accepted}

\titlerunning{The new gravitational lens HE~0230$-$2130}
\authorrunning{L. Wisotzki et al.}

   \maketitle

   \begin{abstract}

We report the discovery of the new gravitational lens system
HE~0230$-$2130, a QSO at redshift $z=2.162$ consisting of at
least five distinct components. Three of these are clearly lensed
images of the QSO, one is most likely the lensing galaxy, while for 
the fifth component the identity is unclear: It could be a fourth
QSO image (if so, then highly reddened), or another intervening 
galaxy, or a superposition of the two. Differential reddening 
seems to be important also for the first three QSO images.
The surface density of faint galaxies near the QSO appears to be
enhanced by a factor of $\ga 2$, indicating the presence of 
a distant cluster close to the line of sight.

      \keywords{%
                Quasars: general --
                Quasars: individual: HE~0230$-$2130 --
                Gravitational lensing
               }
   \end{abstract}

\section{Introduction}

The study of gravitational lenses has become a powerful tool 
to address several distinct cosmological and astrophysical 
questions. These include
the distribution of dark matter in galaxies
(Keeton et al.\ \cite{keeton*98}),
studying dust extinction at redshift $z\gg 0$
(Nadeau et al.\ \cite{nadeau*91}; Jean \& Surdej \cite{jean*98}),
and determining the Hubble parameter $H_0$ from measurements of 
light travel time delay between different lines of sight 
(Refsdal \cite{refsdal64}).
The appeal of this method lies in its complete independence of the
traditional cosmic distance ladder, yielding distances for
high-redshift objects in a single leap. 

The uncertainties of $H_0$ estimation are mainly limited 
by ambiguities in modelling the deflector mass distribution, 
since all relevant \emph{measurement} errors can be reduced 
to insignificance. While `simple' lenses such as double QSOs 
do not strongly constrain the deflector model, 
multiple systems such as quadruply imaged QSOs provide many
additional constraints, 
allowing e.g.\ to independently test the assumed mass 
distribution models (cf.\ Saha \& Williams \cite{saha*97}). 
A problem with quadruple systems, however, can be excessive
symmetry leading to time delays which are short 
compared to intrinsic radio and optical variability timescales,
as in the famous `Einstein Cross' 2237$+$030
or in the `Clover Leaf' H~1413$+$117.

In this paper we present the discovery of 
the new multiple QSO HE~0230$-$2130. The object was originally 
identified as a high-probability QSO candidate in the course 
of the Hamburg/ESO survey (Wisotzki et al.\ \cite{wisotzki*96})
and subsequently observed as part of a large imaging search
for lensed QSOs. The coordinates of this object are
R.A. = 02$^\mathrm{h}$~32$^\mathrm{m}$~33$\fs 1$,
Dec = $-21\degr$~17$'$~26$''$ (J2000.0), as measured in the
\emph{Digitized Sky Survey}.

\begin{figure*}
\begin{picture}(180,44)
\put(0,0){\epsfxsize=4.35cm\epsfbox[98 98 222 222]{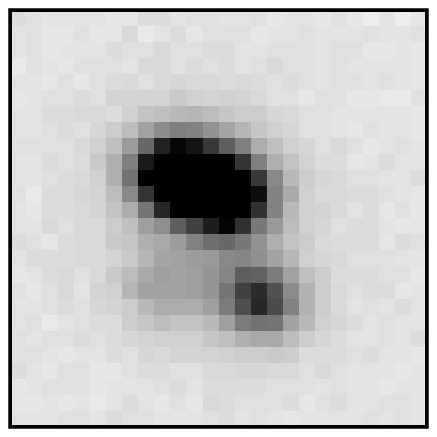}}
\put(4,38){$B$}
\put(45,0){\epsfxsize=4.35cm\epsfbox[98 98 222 222]{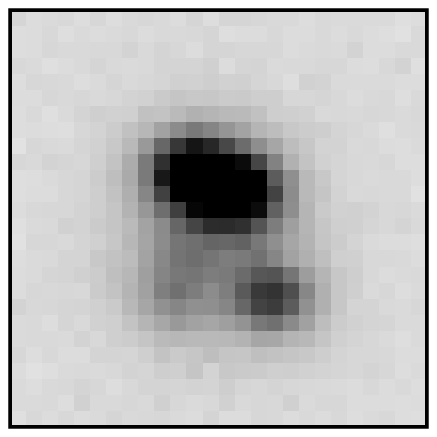}}
\put(49,38){$R$}
\put(90,0){\epsfxsize=4.35cm\epsfbox[98 98 222 222]{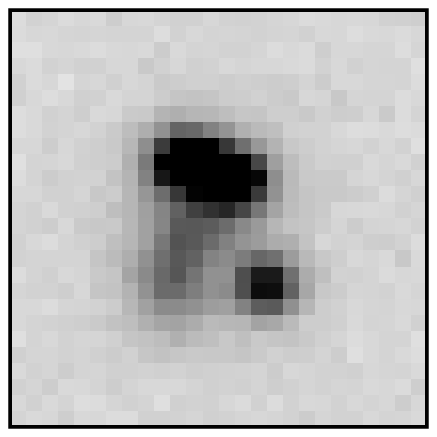}}
\put(94,38){$I$}
\put(135,0){\epsfxsize=4.35cm\epsfbox[98 98 222 222]{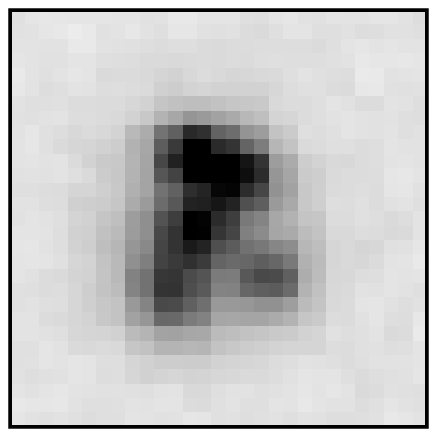}}
\put(139,38){$K_s$}
\end{picture}
\caption[]{\emph{BRIK} band images
           of the multiple QSO system. Each subimage measures 
           $6''\times 6''$; north is up, east is to the left.
           }
   \label{fig:brik}
\end{figure*}

\section{Observations}

\begin{figure}
\epsfxsize=8.8cm\epsfbox[77 87 392 268]{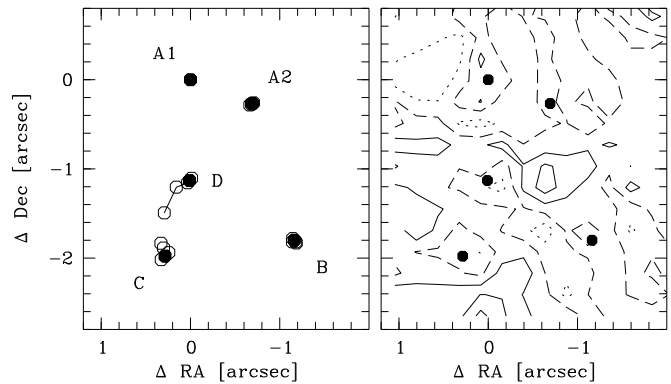}
\caption[]{Left panel: Adopted nomenclature and measured positions 
           of the five components. Filled black circles
           give the positions from Tab.\ \ref{tab:pos}, open circles
           show individual measurements in the different photometric bands.
           Right panel: Contour plot of residual $I$ band image
           after subtraction of five point sources. The dashed contour
           shows the zero level, while negative residuals are dotted,
           and positive residuals are surrounded by the solid contours.
           Contour difference is $2\times$ the photon shot noise
           in each pixel.
           }
   \label{fig:pos}
\end{figure}

Imaging data were obtained at the CTIO 1.5\,m on 13 November 1998,
in Johnson $B$ and $R$, and Cousins $I$. In each band, three images
were taken of 5\,min exposure time each. The seeing was somewhat
variable: $1''$ in $B$, $1\farcs 1$ in $R$, and $0\farcs 8$ in $I$. 
Pixel size was $0\farcs 24$. 
Small sections of these images are reproduced in Fig.\ \ref{fig:brik},
revealing the multiple structure of the QSO already at first glance.
Two almost merging images (A1 and A2; see Fig.\ \ref{fig:pos})
of nearly equal brightness dominate the total magnitudes. 
Another discrete component, B, is clearly apparent with roughly
similar colours. Components C and D are much less prominent, and
in fact appear as two distinct sources rather than just one only
in $I$.

On the same night, near-infrared ($K_s$ band, 2.0--2.3\,$\mu$m) 
images were taken at the CTIO 4\,m telescope with the CIRIM imager.
Total exposure time was 18\,min under $0\farcs 9$ seeing. 
The pixel size of the raw data was $0\farcs 42$, but the single
exposures were combined on a $2\times$ finer grid; the resulting
image is also shown in Fig.\ \ref{fig:brik}. The centrally located
component D is now almost as bright as A1 and A2 and unambiguously
separated from C. A quantitative analysis of the astrometric and
photometric properties is given in the next section.

At the time of these observations, HE~0230$-$2130 had been a mere
QSO candidate based on a digital objective prism spectrum, with
an estimated redshift of $z\simeq 2.165$. Although we had little
doubt about its QSO nature, confirmation was clearly needed. 
We obtained a first low-resolution spectrum on 23 November 1998 
with the ESO/Danish 1.5\,m telescope equipped with DFOSC. 
A slit of $2''$ positioned East-West gave a spectral resolution
of $\sim 20\,$\AA\ FWHM. 
The rather poor seeing of $1\farcs 8$ inhibited all
attempts to separate different components; and the spectrum,
displayed in Fig.\ \ref{fig:spcdanish},
represents more or less the superposition of all components.
It shows a typical QSO with $z = 2.162$, as measured from the 
centroid of the Mg\,{\sc ii} emission line. 

An attempt to obtain individual spectra of different components was
made at the ESO 3.5\,m NTT on 10 December 1998, using the red arm of
EMMI with grism \#3 and a $1''$ slit, at 8\,\AA\ resolution. 
Only one spectrum could be taken, with the slit oriented North-South
crossing A2 and B. At a seeing of $0\farcs 5$, the components were 
well separated, showing two QSO spectra at the same redshifts.
Unfortunately, the spectrum of B suffers from
substantial slit losses: While the flux ratio A2/B is 2.3 in the
red, in good agreement with the PSF photometry described below,
it increases to $\ga 6$ at the blue end around 4000\,\AA .
To correct for these losses, the quotient spectrum was fitted with
a 5th-order polynomial, and the spectrum of B was multiplied by
the fit, thereby adjusting both spectra to the same global level
of relative fluxes. The result is displayed in Fig.\ \ref{fig:spcntt}
and discussed below.

\section{Analysis}

\subsection{Astrometry}

To decompose the complex configuration into discrete sources, 
we employed the DAOPHOT~II software package as implemented within
ESO-MIDAS. For the optical images, a numerical composite PSF was 
built for each image using three bright stars at $\sim 1'$ 
distance to the QSO. In the $K$ band data no bright PSF star 
was available within the small field of view, and a purely analytical
PSF was estimated from the only two isolated stars in the field. 
The QSO was then modelled by the superposition of five point sources, with
positions and PSF scaling factors simultaneously optimised (routine
ALLSTAR). Fitting five components gave always a significantly better fit, 
in terms of residual $\chi^2$, than with four, even in the $B$ band image 
where C and D are both very faint. The fitting results are collected
in Tables \ref{tab:pos} and \ref{tab:phot}. Positions
are measured relative to A1, which was arbitrarily taken as reference point.

For components A1, A2, and B, the positions measured in the four
\emph{BRIK} images are very consistent, and the scatter between the
photometric bands reflects the measurement error. 
For C and D, image positions are consistent only between $I$ and
$K$; at shorter wavelengths, the fitted centroids approach each other
as illustrated in Fig.\ \ref{fig:pos}. This could be
related to different intrinsic colours of the objects, 
but could also be an artefact of low S/N and the
somewhat poorer seeing in the shorter wavelength data.

\subsection{Photometry}

Differential PSF photometry of the QSO components relative to A1 
was available from the ALLSTAR analysis (Tab.\ \ref{tab:phot}).
Flux calibration was established separately using simple aperture 
photometry (aperture diameter was $9\farcs 6$ for \emph{BRI}
and $7\farcs 5$ for $K_s$).
Standard stars from the RU\,149 field (Landolt \cite{landolt*92}) 
served to obtain colour terms and photometric zeropoints
in the optical, while SJ\,9106 (Persson et al.\ \cite{persson*98})
was used for the NIR photometry.

The total magnitudes thus measured are listed in the first line
of Table \ref{tab:phot}, with uncertainty estimates as given in
the DAOPHOT output. We have also determined aperture \emph{BRI}
magnitudes for 14 nearby stars in the field that may be useful 
to serve as reference stars in future monitoring. A list with 
these measurements is available on 
request.\footnote{NDM; email: \texttt{ndmorgan@mit.edu}}

The relative photometry confirms that A1, A2,
and B have very similar optical-NIR colours although B appears 
slightly redder than A1 and A2. C and D are much redder, on the other
hand, so neither can correspond to a single unobscured fourth QSO
image. Because of the apparent positional shift between the bands,
we computed a second model with fixed positions imposed from the 
$I$ band image, fitting only the PSF scaling factors. The resulting
colours are slightly bluer for components B and C, and even much
redder for D. However, inspection of the PSF-subtracted images indicated
that the fit quality of these restricted models was much poorer,
leaving residuals significant on the 2--3$\sigma$ level.

\begin{figure}
\epsfxsize=8.8cm\epsfbox[77 87 341 257]{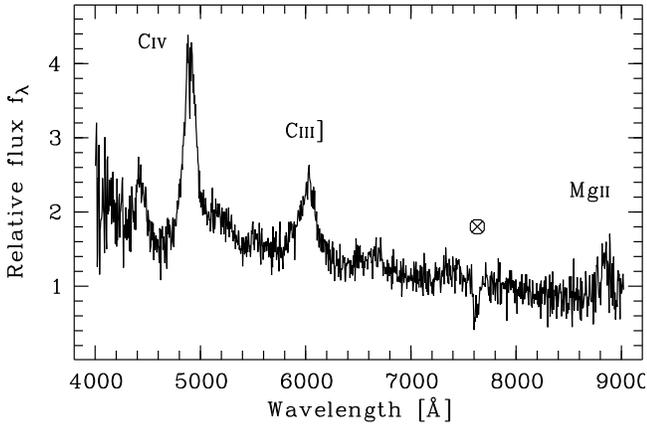}
\caption[]{Spectrum of HE~0230$-$2130 taken with the ESO/Danish
           1.5\,m telescope at poor seeing; 
           the spectrum is the sum of all components.
           }
   \label{fig:spcdanish}
\end{figure}

\begin{table}
\caption{Differential astrometry for HE~0230$-$2130,
         based on the $I$ and $K$ band images.}
\label{tab:pos}
\begin{tabular}{lr@{.}lr@{.}l}\hline\noalign{\smallskip}
Component & \multicolumn{2}{c}{$\Delta\alpha$} & \multicolumn{2}{c}{$\Delta\delta$} \\
          & \multicolumn{2}{c}{[arcsec]} & \multicolumn{2}{c}{[arcsec]} \\
\noalign{\smallskip}\hline\noalign{\smallskip}
A1        &    0&00            &    0&00            \\  
A2        & $-$0&68 $\pm$ 0.01 & $-$0&27 $\pm$ 0.01 \\  
B         & $-$1&17 $\pm$ 0.02 & $-$1&80 $\pm$ 0.02 \\  
C         &    0&28 $\pm$ 0.04 & $-$1&98 $\pm$ 0.04 \\  
D         &    0&01 $\pm$ 0.02 & $-$1&13 $\pm$ 0.03 \\  
      \noalign{\smallskip}\hline
  \end{tabular}
\end{table}

\begin{table}
\caption{Differential photometry of HE~0230$-$2130. The
first row gives the total magnitude from aperture photometry,
all other entries were computed relative to these values. }
\label{tab:phot}
\begin{tabular}{lr@{.}lr@{.}lr@{.}lr@{.}l}\hline\noalign{\smallskip}
Component & \multicolumn{2}{c}{$B$} & \multicolumn{2}{c}{$R$} &
          \multicolumn{2}{c}{$I$} & \multicolumn{2}{c}{$K$} \\
\noalign{\smallskip}\hline\noalign{\smallskip}
Total  &     18&20 &     17&66 &     17&18 &     14&96 \\
       &$\pm 0$&01 &$\pm 0$&01 &$\pm 0$&01 &$\pm 0$&02 \\[0.8ex]
A1     &     19&27 &     18&83 &     18&49 &     16&54 \\
       &$\pm 0$&02 &$\pm 0$&02 &$\pm 0$&02 &$\pm 0$&04 \\[0.8ex]
A2     &     19&27 &     18&89 &     18&43 &     16&65 \\
       &$\pm 0$&02 &$\pm 0$&02 &$\pm 0$&02 &$\pm 0$&04 \\[0.8ex]
B      &     20&06 &     19&45 &     19&00 &     16&95 \\
       &$\pm 0$&02 &$\pm 0$&02 &$\pm 0$&02 &$\pm 0$&04 \\[0.8ex]
C      &     21&87 &     20&42 &     19&74 &     16&80 \\
       &$\pm 0$&08 &$\pm 0$&04 &$\pm 0$&05 &$\pm 0$&04 \\[0.8ex]
D      &     22&02 &     20&62 &     19&63 &     16&65 \\
       &$\pm 0$&24 &$\pm 0$&05 &$\pm 0$&04 &$\pm 0$&05 \\
\noalign{\smallskip}\hline
\end{tabular}
\end{table}

\begin{table}
\caption{Colours of the components in HE~0230$-$2130, based
on the PSF magnitudes from Table \ref{tab:phot}.} 
\label{tab:colours}
\begin{tabular}{lr@{.}lr@{.}lr@{.}l}\hline\noalign{\smallskip}
Component & \multicolumn{2}{c}{$B-R$} & \multicolumn{2}{c}{$R-I$} &
            \multicolumn{2}{c}{$I-K$} \\
\noalign{\smallskip}\hline\noalign{\smallskip}
A1  & 0&44 & 0&34 & 1&95 \\
A2  & 0&38 & 0&45 & 1&78 \\
B   & 0&62 & 0&45 & 2&05 \\
C   & 1&45 & 0&68 & 2&94 \\
D   & 1&40 & 0&98 & 2&98 \\
\noalign{\smallskip}\hline
\end{tabular}
\end{table}

\subsection{Spectroscopic properties}

While in double QSOs there is always the possibility that a true
binary system is being observed, a configuration like that seen in
HE~0230$-$2130 is almost certainly best explained as a lensed system,
even without spectroscopic evidence. Although we do not yet have spectra
of all components, the available data allow nevertheless to confirm
the lens hypothesis beyond all reasonable doubt:

(1) The total spectrum (Fig.\ \ref{fig:spcdanish}) contains no trace 
of absorption features that would be expected if A1 was a star or 
a galaxy. We conclude that A1 and A2 have most probably very similar
spectra, given the broad-band colours.

\begin{figure}
\epsfxsize=8.8cm\epsfbox[77 87 341 257]{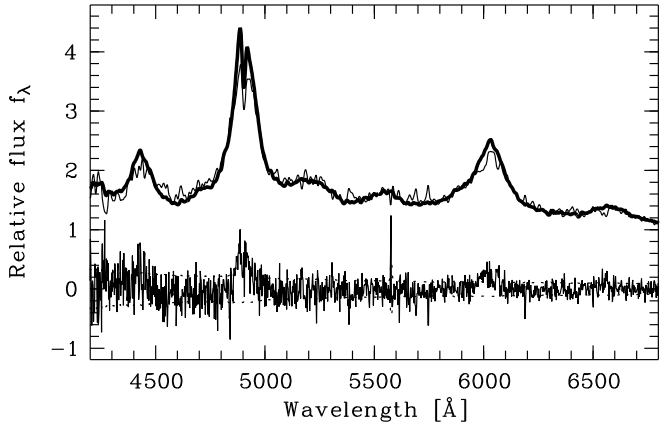}
\caption[]{Comparison between the NTT spectra of components A2 and B
           (slightly smoothed).
           The thick line shows A2, the thin line shows B
           after correction for slit losses (see text).
           Below the unsmoothed difference of both spectra is
           plotted together with the $\pm 1\sigma$ envelope 
           expected from pure shot noise (dashed line).
           }
   \label{fig:spcntt}
\end{figure}

(2) A2 and B have both very similar QSO spectra, apart from the slit
loss effects. Figure \ref{fig:spcntt} shows that the
emission line centroids agree within the measurement accuracy, the 
line widths are equal, and also the strong `associated'
($z_{\mathrm{abs}} \simeq z_{\mathrm{em}}$) C\,{\sc iv} absorption
system is clearly present in both components.

A curious feature, however, are the significant residuals detected in
the difference spectrum A2--B(scaled), indicating non-identical
emission line profiles and/or equivalent widths. Whether this might be
due to differences in intervening line absorption along the lines of
sight, or due to selective continuum microlensing such as proposed in
HE~1104$-$1805 (Wisotzki et al.\ \cite{wisotzki*93}) remains to be
explored; the current data do not permit a more detailed analysis.

\subsection{Nearby galaxies}

\begin{figure}
\begin{picture}(88,44)
\put(0,0){\epsfxsize=4.4cm\epsfbox[48 148 552 652]{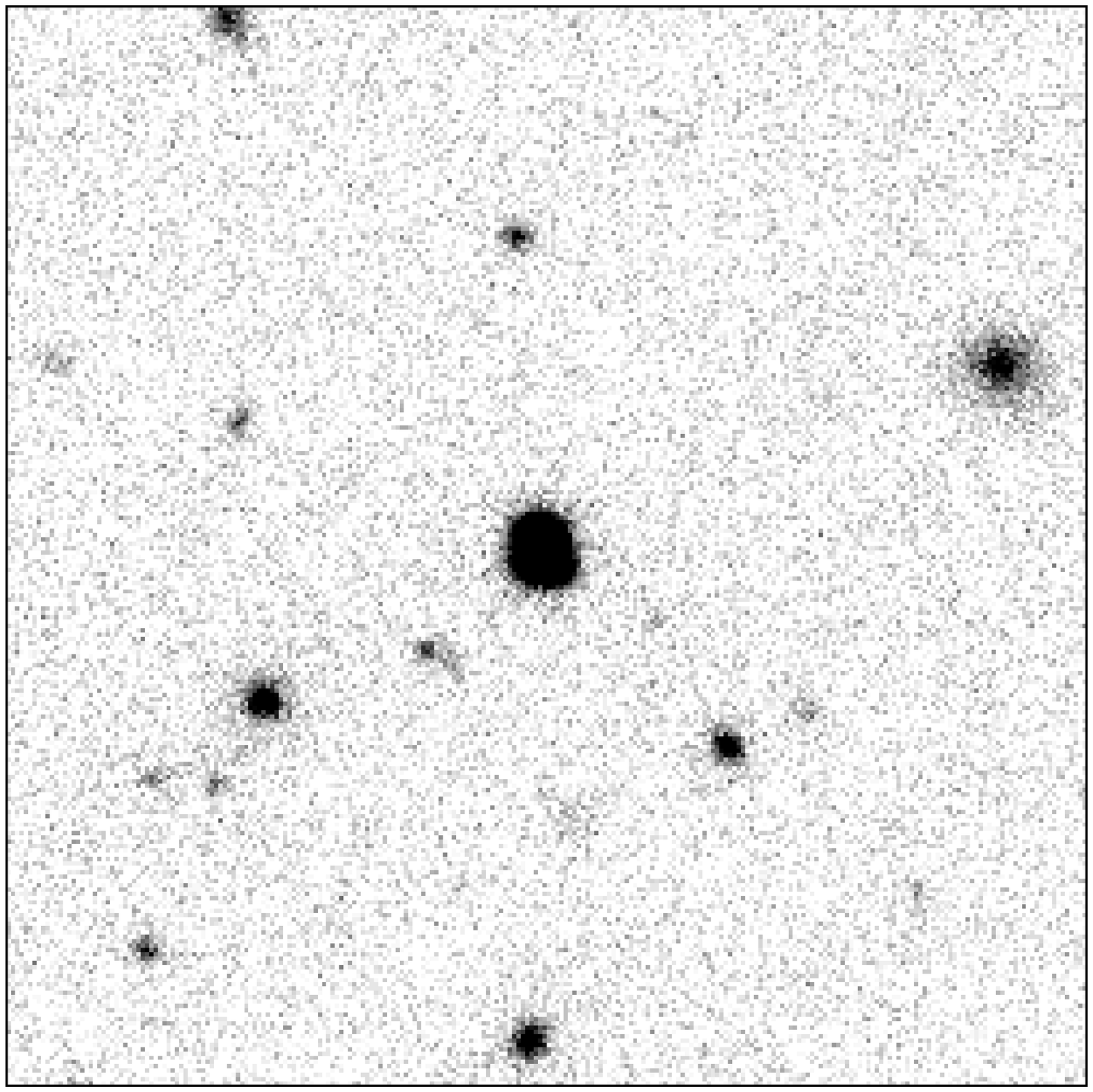}}
\put(3,39){$R$}
\put(44,0){\epsfxsize=4.4cm\epsfbox[48 148 552 652]{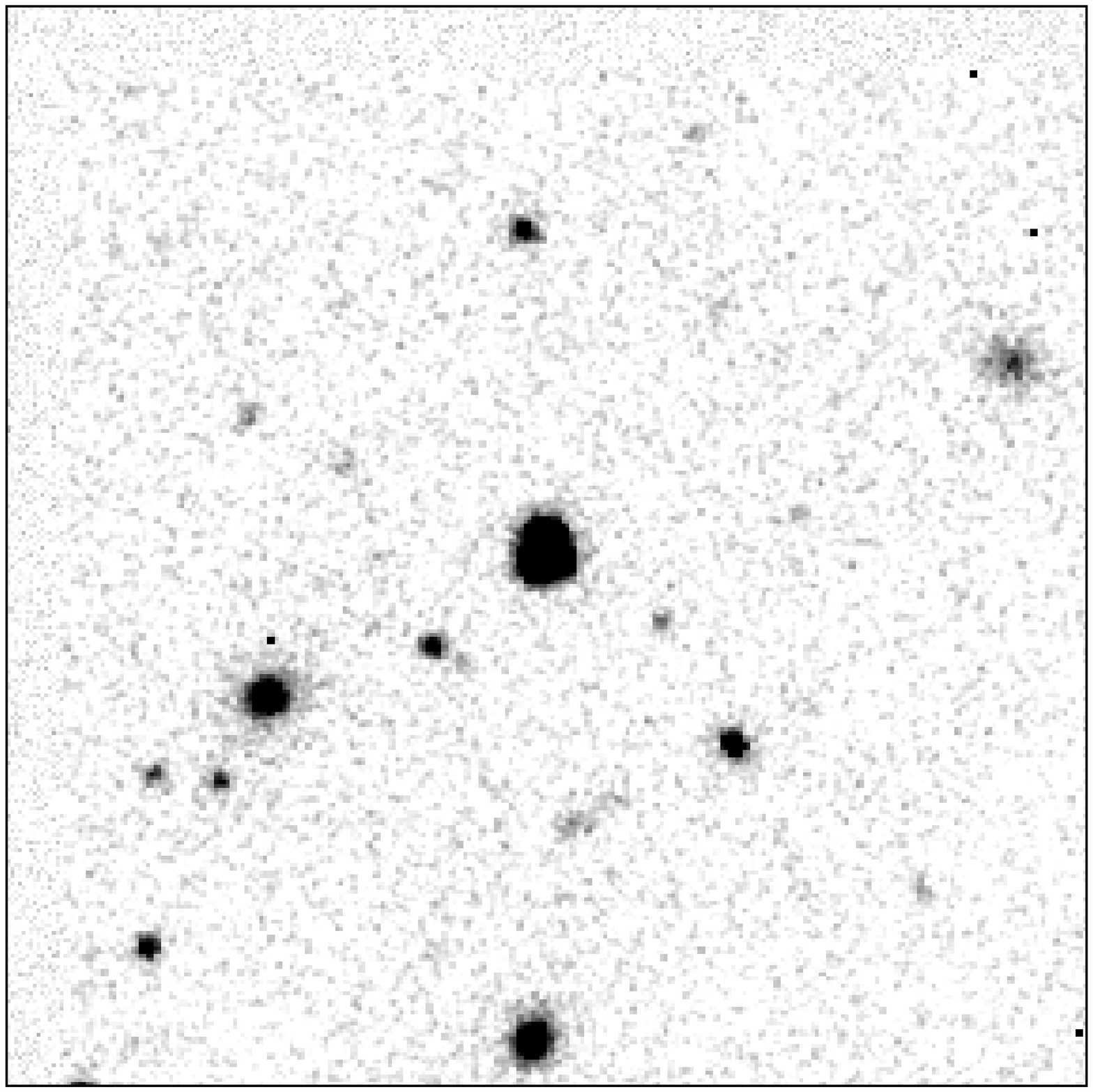}}
\put(47,39){$K_s$}
\end{picture}
\caption[]{$R$ and $K_s$ band images ($1'\times 1'$) showing
the possible galaxy cluster around the QSO. The QSO is the bright object 
in the centre; north is up, east is to the left.
           }
   \label{fig:galenv}
\end{figure}

Visual inspection of the available images 
(see Fig.\ \ref{fig:galenv}) shows a number of faint 
galaxies in the vicinity of the QSO, but most of
these objects are very faint and at the limit of the present data
($R_{\mathrm{lim}}\simeq 23$, $I_{\mathrm{lim}}\simeq 22.5$,
$K_{s,\mathrm{lim}}\simeq 20.5$). The surface density of 
faint objects around the QSO (within $\sim 30''$ radius) 
was found to be enhanced by a factor of 2--3 compared to the 
surrounding regions.
As star-galaxy separation was not reliable for such faint sources,
every detected object was counted, so the overdensity
is a conservative estimate. However, the signal is 
significant only on the 2$\sigma$ level, using Poisson statistics.

Most of the $\sim 20$ objects within $30''$ from the QSO 
are detected in \emph{RI$K_s$} (and expectedly not in $B$).
The $R-I$ and $I-K$ colours show relatively little dispersion
and are consistent with galaxy colours at low to intermediate redshifts,
$z \la 0.6$ (cf.\ Fukugita et al.\ \cite{fukugita*95}).
We conclude that there is some evidence for a cluster near the line 
of sight to HE~0230$-$2130, but that this has to be confirmed by deeper 
imaging and spectroscopy.

\section{Discussion}

Components A1, A2, and B of HE~0230$-$2130 have
been clearly identified, by multicolour imaging and spectroscopically,
as being images of one and the same QSO.
The identities of components C and D have not yet been ascertained.
From their locations, a reasonable guess might be 
that C is the expected fourth image and D the lensing galaxy. 
The colours of D are certainly consistent with this notion,
but the case is more complicated with component C.
Using only positions as constraints, simple models which take 
C to be the 4th image give large residuals, whereas 
models which fit only the positions of images A1, A2 and B 
are poorly constrained.
Moreover, the colours colours of C are very different from 
the confirmed QSO components.
This might indicate either heavy reddening by dust in the fourth image,
or the presence of a second galaxy, or even superposition of a 
galaxy \emph{plus} fourth image. Without spectroscopic information 
on C we presently cannot decide between these possibilities.

If C is partly or largely due to a galaxy, it is \emph{a priori}
most likely (but not granted) that C and D are at the same redshifts,
forming one common deflector plane. 
If, on the other hand, C is formed by just the fourth quasar
image, this would be one of the few known lenses with strongly 
differential dust extinction, and the first optically selected one. 
Furthermore, the inferred amount of differential extinction would be 
$\Delta E(B-V) \simeq 0.4$, just intermediate
between the very small and highly uncertain values found for most 
lenses and the few cases where the light path in the optical
is strongly obscured (cf.\ Falco et al.\ \cite{falco*99}).

HE~0230$-$2130 is an interesting new target for photometric monitoring.
It is moderately symmetric so that the time delay is probably
much shorter than a year, avoiding the considerable 
windowing problem of observations stretching over several years. 
Yet, it displays enough departure from symmetry that the time delay 
will not be uncomfortably low.
In many respects, it resembles PG~1115$+$080, the presently cleanest
system for determination of $H_0$ 
(e.g., Schechter et al.\ \cite{schechter*97}). 
The apparent magnitude and colour of the putative lensing galaxy suggest
that its redshift is not too high, $z_l \la 0.6$.

\begin{acknowledgements}
PLS and NDM gratefully acknowledge support through US NSF grant AST96-16866.
NC acknowledges support from the DFG through grant Re~353/40.
\end{acknowledgements}

\end{document}